\documentclass[11pt]{article}
\usepackage{array}
\usepackage{graphicx}
\usepackage{cite}
\usepackage{textpos}
\usepackage{bbold}

\title{Models of Hidden Purity}
\author{Frank Wilczek  \\
\small\it Center for Theoretical Physics, MIT, Cambridge, MA 02139 USA; \\
\small\it T. D. Lee Institute and Wilczek Quantum Center, \\
\small\it Shanghai Jiao Tong University, Shanghai, China;\\
\small\it Arizona State University, Tempe, AZ, USA; \\
\small\it Stockholm University, Stockholm, Sweden}

\begin{document}

\maketitle

\begin{textblock*}{5cm}(11cm,-8.2cm)
\fbox{\footnotesize MIT-CTP/5260}
\end{textblock*}

\begin{abstract}
I extend, apply, and generalize a model of a quantum radiator proposed by Griffiths to construct models of radiation fields that exhibit high entropy for long periods of time but approach pure states asymptotically.  The models, which are fully consistent with the basic principles of quantum theory, provide coarse-grained models of both realistic physical systems and exotic space-times including black and white holes and baby and prodigal universes.  Their analysis suggests experimental probes of some basic but subtle implications of quantum theory including interference between a particle and its own past, influence of quantum statistical entanglement on entropy flow, and residual entanglement connecting distant radiation with a degenerate source.  
\end{abstract}
\medskip

\bigskip
\bigskip


The relaxation of a quantum system, through radiation, from a highly excited state to its ground state presents an interesting conceptual challenge.    General principles of unitary evolution predict that the resulting radiation field is in a pure state (assuming the ground state is non-degenerate), but general principles of statistical physics suggest that at intermediate times the radiation will exhibit a significant degree of randomness, and in many circumstances will look approximately thermal.   What is the mechanism whereby purity is restored?   What are its observable consequences?  

Here I do two things.  First, I discuss the mechanism of purity restoration precisely and quantitatively in a simple, physically motivated model, for a variety of initial states.  The analysis brings out several interesting effects, including the possibility of interference between a particle and its own past and the influence of quantum statistics on the flow of entanglement entropy.  I also show that ground state degeneracy of the radiating system can lead to residual entanglement with distant radiation, rendering it impure.  The model is capable of many tractable generalizations and suggests practicable experiments.  Second, I will indicate a particular line of generalization that supports coarse-grained but recognizable caricatures of black (and white) hole horizons, and baby (and prodigal) universes.

\bigskip 

{\it Basic Model}\cite{griffiths_book}: We consider first a Hilbert space $S$ spanned by a basis of states $| x \rangle$ with $x$ running through the integers.  
In a unit interval of time the states evolve according to the transformation $V_0$, defined through
\begin{eqnarray}\label{unit_propagation}
| x \rangle ~&\rightarrow&~ | x + 1 \rangle \ \ \ \ \ \ {\rm for} \  x < -1 \ {\rm or} \ x > 0 \nonumber \\
| 0 \rangle ~&\rightarrow&~ \alpha |0\rangle + \beta | 1 \rangle \nonumber \\
|-1\rangle ~&\rightarrow&~ - \beta^* |0 \rangle + \alpha^* |1\rangle 
\end{eqnarray}
with $| \alpha |^2 + |\beta |^2 =1$, so that $\left(\begin{array}{cc}-\beta^* & \alpha \\ \alpha^* & \beta\end{array}\right)$ is a unitary matrix.  We can write Eqn.\,(\ref{unit_propagation}) in matrix form as (the obvious extension of)
\begin{equation}\label{scattering_center}
\left(\begin{array}{cccccc}1 & 0 & 0 & 0 & 0 & 0 \\0 & 1 & 0 & 0 & 0 & 0 \\0 & 0 & -\beta^* & \alpha & 0 & 0 \\0 & 0 & \alpha^* & \beta & 0 & 0 \\0 & 0 & 0 & 0 & 1 & 0 \\0 & 0 & 0 & 0 & 0 & 1\end{array}\right)
\end{equation}
followed by a shift.  

If we view the state-numbers as indicating a linear coordinate, we can regard this model as providing a coarse-grained description of a right-moving particle (e.g., a quantum Hall edge state) interacting with an impurity.  Alternatively, if we view the special state $|0 \rangle$ as the origin in radial coordinates, we can view the positively (respectively, negatively) labelled states as radial coordinates for outgoing (respectively, incoming) radiation that scatters from a center.

Starting with the state $| 0 \rangle$ at time 0, after $n$ steps we have 
\begin{equation} 
V_0^n | 0 \rangle ~=~  \alpha^n | 0 \rangle \, +  \, \beta \alpha^{n-1} |1 \rangle   \, + \, ...     \,+  \, \beta | n \rangle \, .
\end{equation}
This gives a model of an object starting at time $0$ at position $0$ and changing into radiation that propagates away toward larger values of the coordinate.  The process is local and leads to a characteristic exponential decay of amplitude at $0$ together with a specifically structured radiation field, featuring an expanding front with exponential decay toward the origin. 

For our subsequent purposes it is important also to introduce a two-dimensional ``internal'' space $I$ spanned by $| e \rangle, |g \rangle$ and the unitary transformation $W$ on $S \otimes I$ defined through 
\begin{equation}\label{detector}
W ~=~ (\mathbb{1} - |1\rangle \langle 1 | ) \otimes \mathbb{1} + |1\rangle \langle 1 | \otimes \left(\begin{array}{cc}0 & 1 \\1 & 0\end{array}\right)
\end{equation}
Combining these constructions gives us the evolution operator $U$ we want, according to 
\begin{equation}\label{evolver}
U ~=~ WV
\end{equation}
where $V \equiv V_0 \otimes \mathbb{1}$.
This construction can be interpreted as providing a detector that reacts whenever our excitation passes through point 1.   (That is the context in which Griffiths introduced it.)   Here we want to use a slightly different interpretation.   We suppose that our decaying object has an internal degree of freedom corresponding to an excited state $| e \rangle$ and a ground state $| g \rangle$, and this construction is implementing the bookkeeping for that internal degree of freedom.   Alternatively, we might say that the radiating object serves as its own detector.

\begin{figure}[t!]                        
\includegraphics[width=8cm]{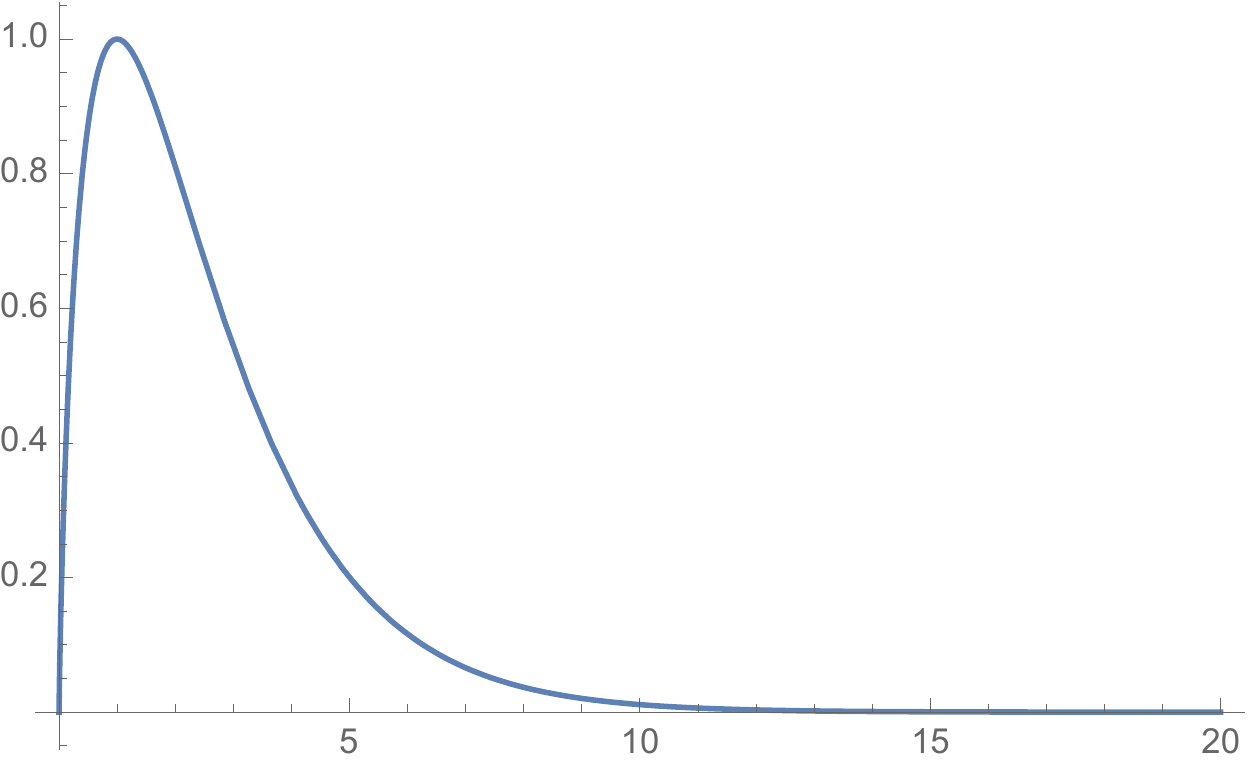}
\caption{The dependence of the entropy of a decaying 2-level system on time in the continuum limit.  The unit of time is the half-life.}
\label{fig:entropy}
\end{figure}

{\it Entropy and Purity Restoration for One-Particle States}:  Beginning at time $0$ with the state
\begin{equation}
\psi_0 (0) ~=~ |0 \rangle \otimes | e \rangle
\end{equation}
then after $n$ time steps we arrive at 
\begin{eqnarray}
\psi_0 (n) ~&\equiv&~ U^n \psi_0 (0)  \nonumber  \\
~&=&~ \alpha^n | 0 \rangle \otimes |e \rangle \, + \, (\beta | n \rangle \, + \, \beta \alpha | n-1 \rangle \, +  \, ... ) \otimes |g \rangle 
\end{eqnarray}
Upon taking the trace over the spatial Hilbert space $S$, we find a simple density matrix for the internal state, viz.
\begin{equation}\label{density0}
\rho_0 (n) ~=~ \left(\begin{array}{cc}|\alpha |^{2n} & 0 \\0 & 1 - |\alpha|^{2n} \end{array}\right)
\end{equation}
and the corresponding von Neumann entropy
\begin{equation}
{\rm Ent}_0 (n) ~=~ - (|\alpha | ^{2n} \log_2 | \alpha| ^{2n} \, + \,  (1 - |\alpha | ^{2n}) \log_2 (1 - |\alpha | ^{2n}) )
\end{equation}
The entropy vanishes at time $n= 0$ and also asymptotically as $n\rightarrow \infty$.  (Here and below ``entropy'' generally will refer to entanglement entropy.  Ref. \cite{popescu} is an insightful discussion of the relationship between entanglement and thermodynamic entropies.) This corresponds, of course, to the fact that the decaying object begins in a pure state - the excited state - and approaches another pure state - the ground state.   (Page \cite{page_curve} has emphasized the interest of such curves in the context of black hole physics.) In between the entropy reaches a maximum value near 1, at a time corresponding to the half-life of the object.  In the continuum limit we find a universal form for the time-dependent entropy, viz.
\begin{equation}
{\rm CEnt} (t) ~=~ t\,  2^{-t} - (1- 2^{-t}) \log_2 (1 - 2^{-t})
\end{equation}
where the unit of time is the half-life. Figure\,\ref{fig:entropy} displays this function.

The essence of the purity restoration mechanism is simply to follow out the full temporal course of emission from an excited state.  Superposing many different emitters of this general sort (see below), the radiation field can come to look complex or random to most local observables, but the underlying mechanism of purity restoration still bears out. 

The radiation field evolves toward a limiting pure state 
\begin{equation}
\beta ( |M \rangle \, + \, \alpha |M-1 \rangle \, + \, \alpha^2 | M-2 \rangle \, + \, ... )
\end{equation}
with $M \rightarrow \infty$ and $M$ increasing by unity through each time step.  At any finite time, the series is truncated when it reaches $|1 \rangle$, and we have a state whose norm is less than unity.  The probability deficit is made up by the possible absence of radiation, i.e., the probability that the object has not yet decayed.   The correlations in the radiation field which enforce its purity have observable consequences.  To bring them out, we should produce interference between different parts of the pure state wave function.   This can be done by inserting detectors in appropriate ways, that depend on space and time, to take samples of the wave function and then cause those samples to interfere.  In an optical context, these operations could be implemented effectively using mirrors and beamsplitters.   In this way, the radiated particle can be made to interfere with its own (virtual) past or future.  Analogous emission of slower quanta, e.g. polaritons, from engineered sources might allow detailed study more easily.  

\begin{figure}[t!]                        
\includegraphics[width=8cm]{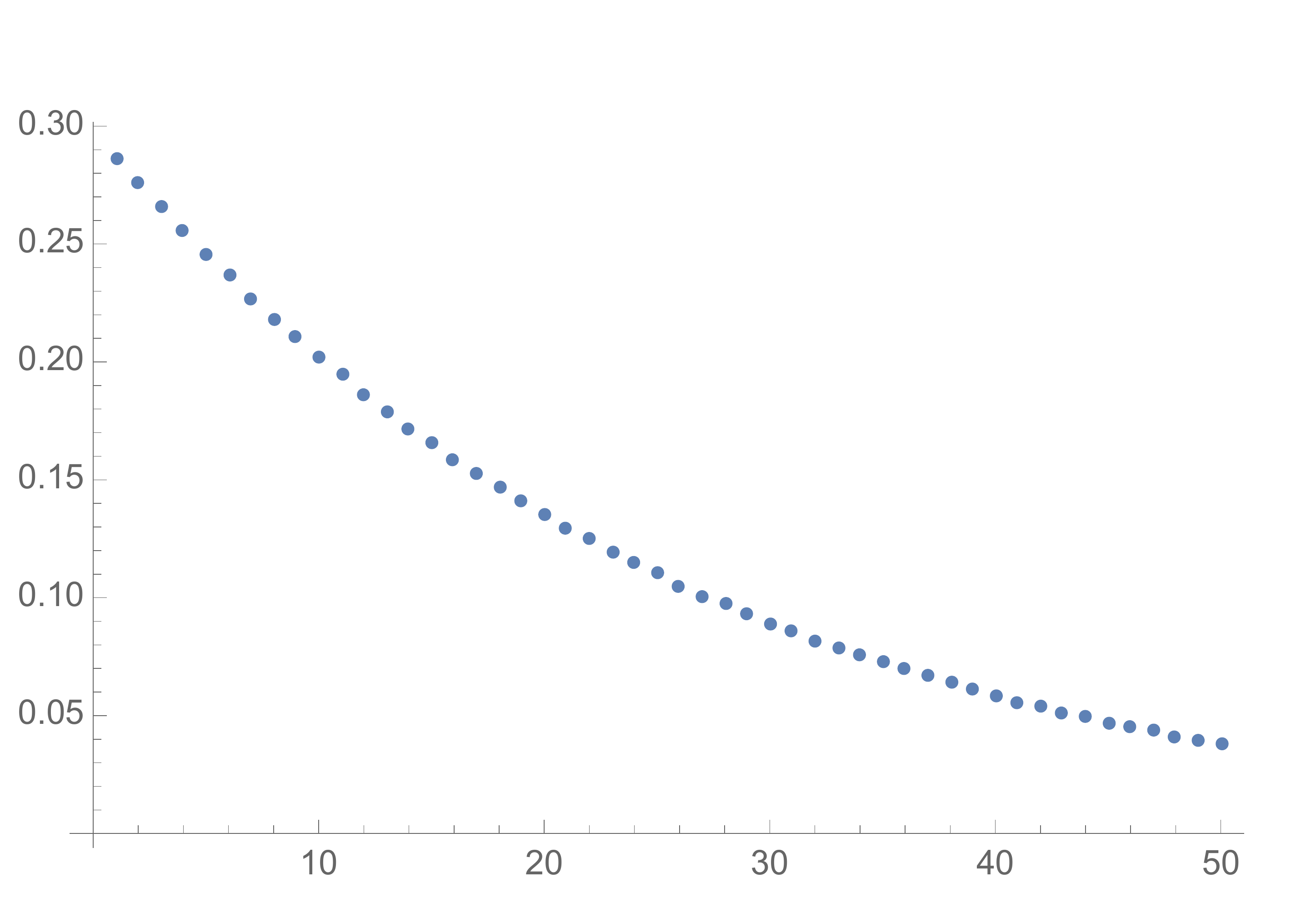}
\caption{Entropy of a radiation field including both scattering and capture, from Eqn.\,(\ref{shadow_density}) with $|\alpha|^2 = .95$.}
\label{fig:shadow_entropy}
\end{figure}

It is also interesting to consider the initial state 
\begin{equation}
\psi_{-1} (0) ~=~ |-1 \rangle \otimes | e \rangle
\end{equation}
In this case, we find the density matrix, corresponding to Eqn.\,(\ref{density0})
\begin{equation}\label{shadow_density}
\rho_{-1} (n) ~=~ \left(\begin{array}{cc}|\alpha |^{2n-2} - |\alpha |^{2n}& 0 \\0 & 1 - |\alpha|^{2n-2} + |\alpha |^{2n} \end{array}\right)
\end{equation}
The corresponding entropy, for a representative value of the parameter, is displayed in Figure\,\ref{fig:shadow_entropy}. 
The difference between $\rho_0 (n) $ and $\rho_{-1} (n) $ arises because the latter includes, at step $n=1$, not only the particle ``captured'' at the origin (with amplitude $-\beta^*$) but also the particle missing the target and ``propagated through'' to position $+1$ (with amplitude $\alpha^*$).  Given the radiation process, unitarity implies the presence of the latter contribution.

{\it Entropy and Purity Restoration for Simple Two-Particle States}:  The problem of calculating the radiation field generally becomes more difficult once we allow the radiators to interact with one another. Interesting phenomena arise already when we consider the effect of quantum statistics.  Then we must work in the symmetric or antisymmetric tensor products, and the calculations get more complicated, reflecting important quantitative effects of entanglement.  

The simplest non-trivial case arises when we evolve appropriately symmetrized products $\psi_0 (0) \otimes \psi_{-1} (0)$.  A striking and at first sight surprising result is that the entropy emerging from this two-particle fermionic state is precisely equal to that arising from the one-particle state $\psi_0 (0)$.  This happens because the states $|-1 \rangle, |0 \rangle$ both evolve into linear combinations of $|0\rangle, |1\rangle$, so that the initial antisymmetric product must evolve into the transported antisymmetric product, with unit amplitude (up to a phase).  After that, the descendants of $|1\rangle$ evolve trivially and stay ahead of the descendants of $|0 \rangle$, so all the entropy is associated with the latter.  For similar reasons, the evolution of antisymmetric products involving solid blocks $|-k\rangle,  |-k-1\rangle, ..., |-k-m\rangle$ of fermionic occupation essentially reduces to single-particle dynamics - a phenomenon reminiscent of Newton's cradle.  

\begin{figure}[t!]                        
\includegraphics[width=8cm]{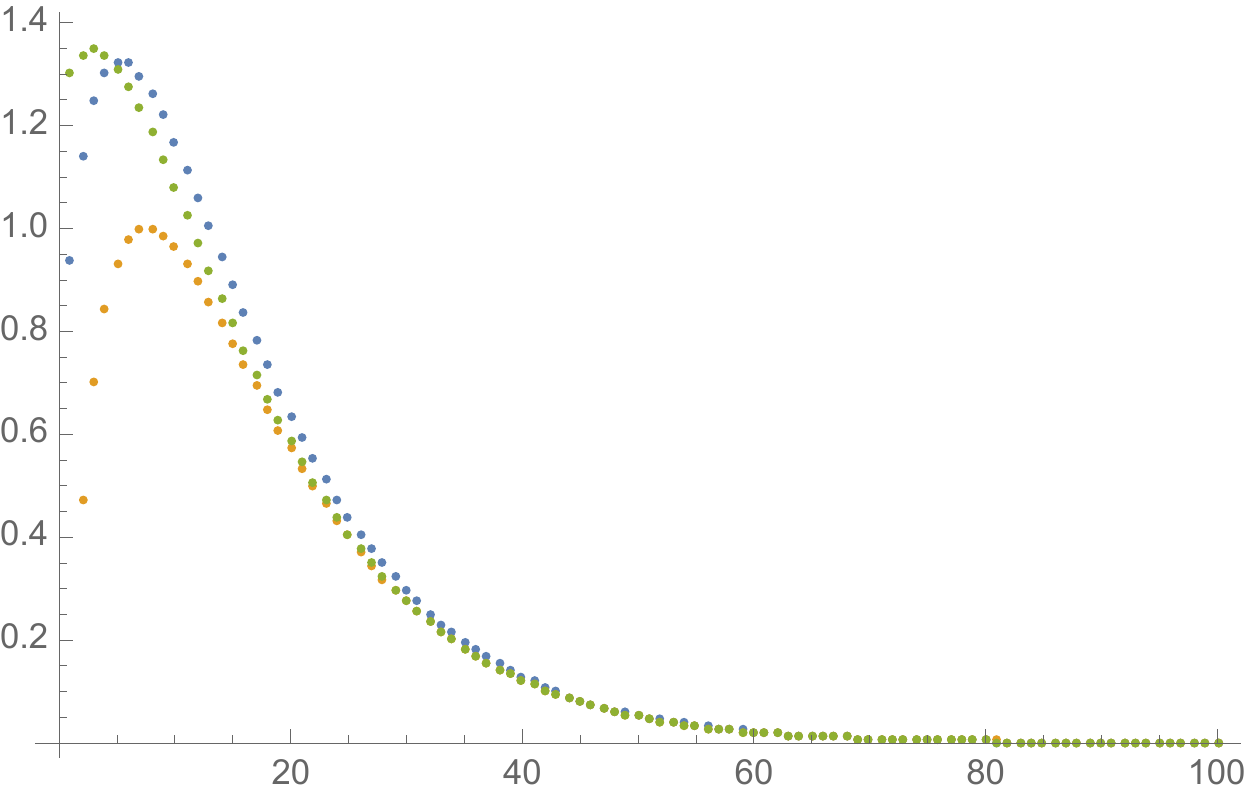}
\caption{Entropy of the radiation field evolving from initial fermionic (orange), bosonic (green), and distinguishable (blue) initial states based on $\psi_0 (0) \otimes \psi_{-1} (0)$.  }
\label{fig:statistics_comparison}
\end{figure}

The bosonic calculation is more intricate.   A straightforward but lengthy calculation gives us, after $n$ time steps, a diagonal density matrix with three entries, corresponding to $|ee\rangle \langle ee|, \frac{1}{2} (|eg\rangle + |ge\rangle) (\langle eg | + \langle ge |), |gg \rangle \langle gg|$, having values $2(1-q)q^{2n-1}, 4 q^n (1- q^{n-1})(1-q) + (2q-1)^2 q^{n-1}, 2q(1-q)(1-q^{n-1})^2 + (2q-1)^2 (1-q^{n-1}) + 2q(1-q)$, where $q \equiv |\alpha|^2$.    Figure\,\ref{fig:statistics_comparison} displays the distinguishable, fermionic, and bosonic entropies for $q= .9$.   The characteristic bunching of bosons and anti-bunching of fermions is there clearly visible.  

{\it Model of Residual Entanglement}: Essentially new phenomena arise if we move the internal degree of freedom from $|1 \rangle$ to $|0 \rangle$, by replacing the $W$ of Eqn.\,(\ref{detector}) with
\begin{equation}\label{moved_detector}
\tilde W ~=~ (\mathbb{1} - |0\rangle \langle 0 | ) \otimes \mathbb{1} + |0\rangle \langle 0 | \otimes \left(\begin{array}{cc}0 & 1 \\1 & 0\end{array}\right)
\end{equation}
and evolve using $\tilde U = \tilde W V$.   Now ${\tilde \psi}_0 (0) = |0 \rangle \otimes |e \rangle$ evolves into
\begin{eqnarray}
{\tilde \psi}_0 (n) ~&\equiv&~ {\tilde U}^n {\tilde \psi}_0 (0)  \nonumber  \\
~&=&~  \alpha^n |0\rangle \otimes |v(n)\rangle + \beta \sum\limits_{j=1}^{n} \alpha^{n-j} | j \rangle \otimes |v(n-j) \rangle  
\end{eqnarray}
where
\begin{equation}
| v(n) \rangle ~\equiv~ \frac{1}{2} (1 + (-1)^n) |e\rangle \, + \, \frac{1}{2} (1 - (-1)^n) |g\rangle
\end{equation}
Tracing over the radiation field, we find that the source density matrix is 
\begin{eqnarray}
{\tilde \rho}_0 (n) ~&=&~ \frac{1+ q^{n+1}}{1+q}  |e\rangle \langle e| \, + \, \frac{q-q^{n+1}}{1+q}|g\rangle \langle g| \ \ \ \ \ {\rm (n \ even)} \nonumber \\
                            ~&=&~ \frac{1-q^n}{1+q}   |e\rangle \langle e| \, + \,  \frac{q + q^n}{1+q}  |g\rangle \langle g|  \ \ \ \ \ {\rm (n \ odd)} \nonumber \\
                            ~&\rightarrow&~ \frac{1}{1+q} |e\rangle \langle e| \, + \, \frac{q}{1+q}  |g\rangle \langle g|  \ \ \ \ \ n \rightarrow \infty
\end{eqnarray}
Thus, even as $n \rightarrow \infty$ and the radiation physically departs from the source, there is residual entropy, indicating residual entanglement between source and radiation.

Tracing instead over the source, we find that the radiation field is a mixture of two propagating pure states, an ``even'' state supported on space-time points with even parity for the space coordinate minus the time coordinate, and an ``odd'' state supported on space-time points with odd parity for the space coordinate minus the time coordinate.   Radiation samples taken from within either state will interfere with one another, while radiation samples from the distinct states will not interfere with one another.   If we regard the operator defined by Eqn.\,(\ref{moved_detector}) as implementing a detector, we might interpret this situation verbally by saying that the detector has measured the moment of emission, modulo 2, and thereby collapsed the wave-function for the radiation into two separate, non-interfering components.  Alternatively, if we regard that operator as implementing an internal degree of freedom, we might interpret the situation as revealing the existence of alternative final (ground) states for the emitter.   Either interpretation can be appropriate, within different applications of models of this kind.  For example, if we want to model the influence of complex environments on radiation processes, we can include multiple detectors at different times and places, stochastically distributed, and calculate how they introduce entropy and decoherence into the radiation.   On the other hand, if we want to use subtle aspects of the radiation field to diagnose subtle aspects of a radiating system, we can strive to eliminate inadvertent disturbances and then explore the hidden (de)coherence of the radiation through appropriate interference experiments.  

It is enlightening to introduce, as a formal device, asymptotic wave functions and density matrices, according to the template $$\phi_\infty (\xi) \, \equiv \, \lim\limits_{t\rightarrow \infty} \phi(t, x = t- \xi) $$
In this way, we can define the aforementioned parity cleanly as a symmetry of the asymptotic wave function and density matrix.

{\it Generalizations}: This model can be extended in many ways whilst remaining reasonably tractable.  The simplest is to consider a large number $N$ of independent radiators, each communicating with an independent mode of radiation.  This can be implemented by taking the tensor product of $N$ copies of the basic model. If the half-lives are all equal, we simply multiply the entropy function by $N$.   We can, of course, allow a spectrum of half-lives. We can also allow the particles to start at different states $\psi_{-k} (0) \equiv |-k \rangle \otimes | e \rangle, k \geq 1$, which will simply delay the fields and entropy calculated for $\psi_{-1}$ by $k-1$ units of time.  As long as all the particles are independent and distinguishable, the entropies will be additive.  In this way we can build up radiation fields that have very high entropy for long periods of time, yet evolve into pure states.  An especially interesting generalization is to include dynamical loops, as I will now discuss.


{\it Loops, Long-lived Centers, Black Holes and Baby Universes}: The problem of hidden purity arises acutely in the theoretical description of black holes as quantum objects \cite{polchinski} \cite{bh_recent_review}.   One can certainly imagine creating a black hole through gravitational collapse of matter in a pure quantum state.  On the other hand, Hawking's famous calculations indicate that black holes can emit radiation with high entropy content.  

Black holes in general are very complex objects, and their adequate description lies far beyond the simple ``Tick-Tock'' (i.e., discrete space, discrete time, unitary evolution) models considered here.  But enriched versions of Griffiths' original model bring in dynamical space-time geometry, and provides more recognizable caricatures featuring many internal degrees of freedom, long time delays, and ``one-way'' processes. (An alternative method of bringing in long time delays, which more closely resembles time dilation, will be described elsewhere.)  

\begin{figure}[t!]                        
\includegraphics[width=8cm]{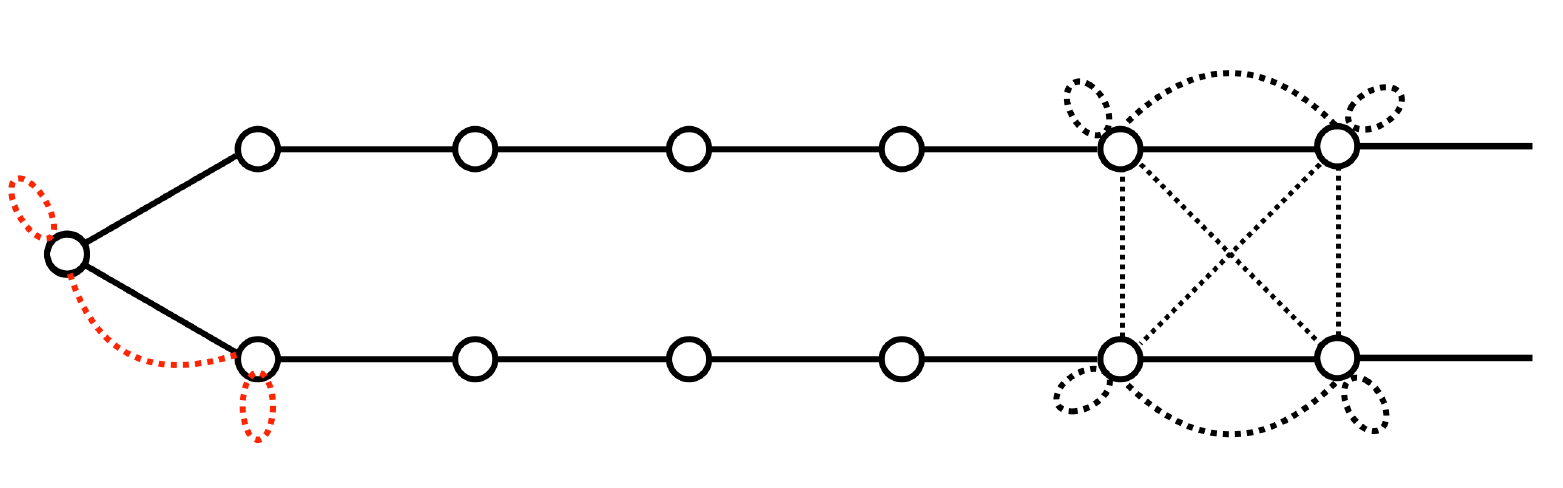}
\caption{State diagram for the models in the text.   The red dotted lines indicate possible non-trivial transition elements in the simple model, Eqn.\,(\ref{scattering_center}).   The black dotted lines indicate possible non-trivial transition elements as indicated by the * entries in Eqn.\,(\ref{pinch_operator}).   In each case, the basic unitary evolution is built up from those indicated matrix followed by a shift.  (Strictly speaking, one should draw directed links.)  For suitable choices of the * entries we can cordon off spaces that are minimally connected or disconnected from the asymptotic regions.   The location of the pinch and its severity can vary with time exogenously or endogenously, as described in the text.}
\label{fig:spacetime_pinch}
\end{figure}

The key idea is to replace Eqn.\,(\ref{scattering_center}) with a more flexible form
\begin{equation}\label{pinch_operator}
\left(\begin{array}{ccccccc} 
\mathbb{1} &  & &  &  &  &    \\ 
& *& * &  & *& *  & \\ 
 & *& * &  & *& *  & \\ 
&  &  & \mathbb{1}_{l}  &  & & \\
& *& * &  & *& *  & \\ 
& *& * &  & *& *  & \\ 
 &  & &  &  &  & \mathbb{1}
\end{array}\right)
\end{equation}
where the $*$ entries taken on their own define a $4\times4$ unitary matrix, the $\mathbb{1}$ are infinite identity matrices, $\mathbb{1}_l$ is an $l\times l$ identity matrix, and all blank entries vanish.   Figure \ref{fig:spacetime_pinch} depicts the effect of Eqn.\,(\ref{pinch_operator}).

Let us denote the $4\times 4$ unitary matrix in Eqn.\,(\ref{pinch_operator}) by $K$.  When
\begin{equation}
K ~=~ \left(\begin{array}{cccc} 
1& 0 & 0&  0    \\ 
0 & 0&  1& 0   \\ 
0  &1 & 0 &  0 \\ 
0 &  0 & 0  & 1 
\end{array}\right)
\end{equation}
the dynamics takes place in two completely separate systems: a circle of length $l$ behind the pinch point, and an infinite line outside it.  When $K$ is approximately of this form, we have an inner region that supports many internal states (i.e., the positions of particles inside) that decay only very slowly, for large $l$, into its (empty) ``ground state''.   This provides a model of radiation from a complex molecule, liquid drop nucleus or black hole.  In the latter case, the pinch point would be interpreted as the horizon.  

We can allow our unit-step unitary evolution operator to vary with step number.  It can differ in the numerical structure of $K$, the size of $l$, or the position of the pinch point.  It can differ in its position, the numerical structure of $K$, and the value of $l$.  In this way, we can allow an effective horizon to form, fluctuate, or disappear.   The horizon can also pinch off completely and permanently, giving birth to a baby universe.    We can also implement the time-reversal of that process, whereby a prodigal universe comes into contact with a hitherto separate universe, or rejoins it after a long hiatus.   Let me emphasize again that all these possibilities, by construction, are fully consistent with the conventional framework of quantum theory and its probability interpretation.   (Related constructions have been implemented using continuous changes in the boundary conditions of wave functions \cite{balachandran} \cite{topology_change}.)  The creation and destruction of practically inaccessible baby and prodigal universes could supply a mechanism for irreducible decoherence within the accessible universe.  

The parameters governing the unit-step evolution operators can be imposed exogenously, for some modeling purposes, but might also be considered as defining a dynamical system in their own right, possibly influenced by the state of the system.   For example, one can model the (approximate) one-way character of black holes absorption by allowing the pinch to open up when a particle is incident at the horizon, then close back up after the particle crosses the horizon.  This can be implemented in a unitary way, using a detector construction analogous to Eqn.\,(\ref{detector}).   It involves a kind of Maxwell demon, and one could consider the implications of that demon having limited memory capacity \cite{bennett}.   

In conclusion, let me remark that Tick-Tock models correspond directly to programs for sequential quantum computers.

{\it Acknowledgement}: I would like to thank Wu Biao, Jordan Cotler, Chirag Falor, Martin Greiter, Hans Hansson, Sid Morampudi, Alfred Shapere and George Zahariade for helpful comments. This work is supported by the U.S. Department of Energy under grant Contract  Number DE-SC0012567, by the European 
Research Council under grant 742104, and by the Swedish Research Council under Contract No. 335-2014-7424.

\end{document}